\documentclass[]{spie}  
\usepackage[]{graphicx}
\usepackage{float}

\title{Design of the iLocater Acquisition Camera \\
Demonstration System} 

\author{ Andrew Bechter*\supit{a}, Jonathan Crass\supit{a}, Ryan Ketterer\supit{a}, Justin R. Crepp\supit{a}, David King\supit{b}, Bo Zhao\supit{c}, Robert Reynolds\supit{d}, Philip Hinz\supit{e}, Jack Brooks\supit{a}, Eric Bechter\supit{a}
\skiplinehalf
\supit{a}Department of Physics, University of Notre Dame, 225 Nieuwland Science Hall, Notre Dame, IN 46556, USA; \\
\supit{b}Institute of Astronomy, University of Cambridge, Cambridge CB3 0HA, United Kingdom;\\
\supit{c}Department of Astronomy, University of Florida, Gainesville, FL 32611;\\
\supit{d}Large Binocular Telescope Observatory, University Of Arizona, 933 N. Cherry Ave., Tucson, AZ 85721;\\
\supit{e}Department of Astronomy/Steward Observatory, University of Arizona, 933 North Cherry Avenue, Tucson, USA
}

\begin{document} 
  \maketitle 

\begin{abstract}
Existing planet-finding spectrometers are limited by systematic errors that result from their seeing-limited design. Of particular concern is the use of multi-mode fibers (MMFs), which introduce modal noise and accept significant amounts of background radiation from the sky. We present the design of a single-mode fiber-based acquisition camera for a diffraction-limited spectrometer named ``iLocater." By using the ``extreme" adaptive optics (AO) system of the Large Binocular Telescope (LBT), iLocater will overcome the limitations that prevent Doppler instruments from reaching their full potential, allowing precise radial velocity (RV) measurements of terrestrial planets around nearby bright stars. The instrument presented in this paper, which we refer to as the acquisition camera ``demonstration system," will measure on-sky single-mode fiber (SMF) coupling efficiency using one of the 8.4m primaries of the LBT in fall 2015. 

\end{abstract}



\section{INTRODUCTION}
\label{sec:intro}
Current radial velocity (RV) spectrometers have reached a limit of 1 m/s single measurement precision due to systematic effects, many of which result from their seeing-limited design. In order to push the capability of planet finding spectrometers to detect Earth-size worlds located in the habitable zone, this level of performance needs to be improved by nearly an order of magnitude\cite{Crepp}. A promising solution for achieving sub meter-per-second precision involves the use of a highly stable, diffraction-limited spectrometer illuminated by a single-mode fiber (SMF). In order to realize this solution, the incident optical beam delivered from a ground-based telescope must be corrected by an adaptive optics (AO) system for efficient fiber coupling. 

Diffraction-limited RV spectrometers offer a number of design advantages compared to their seeing-limited counterparts\cite{Schwab}. In addition to lending themselves to high spectral resolution, diffraction-limited spectrometers achieve a compact optical design that translates into improved stability, higher optical quality, and lower cost\cite{Robertson}. Single-mode fiber-fed spectrometers are also naturally compatible with laser frequency comb (LFC) calibration sources at near infrared (NIR) wavelengths\cite{LFC-SMF}. While implementing an SMF injection system into RV spectrograph design presents several technical challenges, overcoming these will enable the pursuit of new exoplanet science cases.



\subsection{What is iLocater?} 
\label{sec:What is iLocater}

The \textbf{i}nfrared \textbf{L}arge Bin\textbf{oc}ul\textbf{a}r \textbf{T}elescope \textbf{e}xoplanet \textbf{r}econnaissance (``iLocater") spectrograph is a new instrument being built at the University of Notre Dame to exploit the advantages of working at the diffraction limit. This ultra-precise ($\Delta$RV = 20 cm/s), high-resolution (R=150,000), diffraction-limited Doppler spectrograph will detect small RV semi-amplitude signals around bright natural guide stars with ``extreme" AO correction. iLocater has been officially approved for installation at the Large Binocular Telescope (LBT), and will be commissioned on a timescale commensurate with the launch of NASA's Transiting Exoplanet Survey Satellite (TESS). In addition to detecting Earth-like planets in the habitable zone around low-mass stars, iLocater will acquire spin-orbit measurements of \textit{terrestrial} planets, conduct the first systematic study of planets in close-separation binaries, and enable many other cutting-edge exoplanet (and non-exoplanet) science cases.\footnote{iLocater.nd.edu}

iLocater consists of three major instrumentation deliverables: the described spectrograph, an acquisition camera (AC) system which couples light from each telescope dish to separate SMFs, and a wavelength calibration unit (fiber Fabry-P\'erot etalon\cite{Gurevich} and Laser Frequency Comb LFC\cite{MenloSystems}) stable to $< 10$ cm/s precision. The spectrograph and calibration source will be installed in a thermally and vibrationally isolated instrument room located inside the telescope pier of the LBT. The acquisition camera system will ultimately be installed at a newly created central instrument port of the Large Binocular Telescope Interferometer \cite{Hinz} (LBTI) capable of receiving light from both telescope dishes. At this location, the AC will benefit from an extremely well-corrected beam of starlight from LBT's dual deformable secondaries and LBTI's pyramid wavefront sensor system.\cite{Esposito}.

The most difficult challenge in achieving iLocater's diffraction-limited performance is the efficient coupling of starlight into SMFs\cite{Jovanovic}, as they are an order of magnitude smaller than multi-mode fibers (MMFs) commonly used in seeing-limited spectrographs. However, AO correction permits the use of SMFs by providing an order of magnitude improvement in spatial resolution. As there has yet to be an unambiguous demonstration of efficient, on-sky SMF coupling, we have designed and are constructing a ``demonstration system" that will inform the design of the final iLocater instrument by measuring SMF coupling efficiencies on bright target stars. This paper describes the optical and mechanical design of the acquisition camera ``demonstration system," which will be installed and tested using an existing mounting structure\cite{SHARK} on the LBT in fall 2015. 
 
\subsection{Modal Noise} 
\label{sec:MODAL NOISE} 
Single-mode fibers can theoretically eliminate a significant source of noise in RV spectrometers, colloquially referred to as ``modal noise" - a phenomenon introduced by any optical fiber that can support more than one guided mode (i.e. a multi-mode fiber). When this effect occurs, the modal power distribution (MPD) presents itself as a speckle pattern in the far-field output of the fiber, due to the interference of the electric-field patterns of the individual fiber modes\cite{Wood}. Instability of the injected beam or perturbations of the optical fiber (from mechanical stresses) change the electric field boundary conditions within the fiber causing temporal variations in the speckle pattern. The resulting unpredictable speckle pattern with a temporally shifting intensity distribution is known as modal noise. Throughout an exposure with a spectrograph, modal noise can be confused with a shifting spectral line of a star and thus effectively limits the precision of RV measurements \cite{Jovanovic}. 

For high-resolution spectrographs, light at the detector becomes effectively coherent such that the signal-to-noise ratio (SNR) is no longer photon-limited, but modal noise-limited\cite{Lemke}. This makes modal noise one of the most important sources of error to mitigate for RV precision at the cm/s level. Furthermore, since fewer modes are excited in fibers at NIR compared to optical wavelengths, the effect of modal noise at NIR wavelengths causes a larger intensity shift in the MPD, making NIR fiber-coupled instruments more susceptible to the uncertainty induced by modal noise\cite{McCoy}. Despite the decrease in precision, current generation Doppler spectrometers use MMFs as the preferred optical fiber type, as their large core diameter and low wavefront tolerances allow for higher throughput using seeing-limited incident beam.

The optical properties of the fiber used by iLocater only supports the LP$_{01}$ fundamental mode (TEM$_{00}$ mode). This single-mode operation allows for an ultra stable intensity distribution at the fiber output with a radial mode shape approximated by a Gaussian function\cite{Wagner}. This distribution is independent of the injected beam stability at the fiber face and from mechanical stresses. Such a highly stable output distribution is ideal for the extraction of precise RV measurements using high resolution spectrographs. 

\section{OPTICAL DESIGN} 
\label{sec:OPTICAL DESIGN} 
The iLocater demonstration system consists of an acquisition camera (AC) and an optical power monitoring system designed for installation on the DX (right) side of the LBT located just prior to the universal beam combiner (UBC) in LBTI\cite{Hinz}. The AC is designed principally for SMF coupling in the Y-band (950-1120nm); all other ancillary functions are designed to assist in optimizing coupling efficiency and maintaining stable alignment. 

\begin{figure}[H]
   \begin{center}
   \begin{tabular}{c}
   \includegraphics[height=9cm]{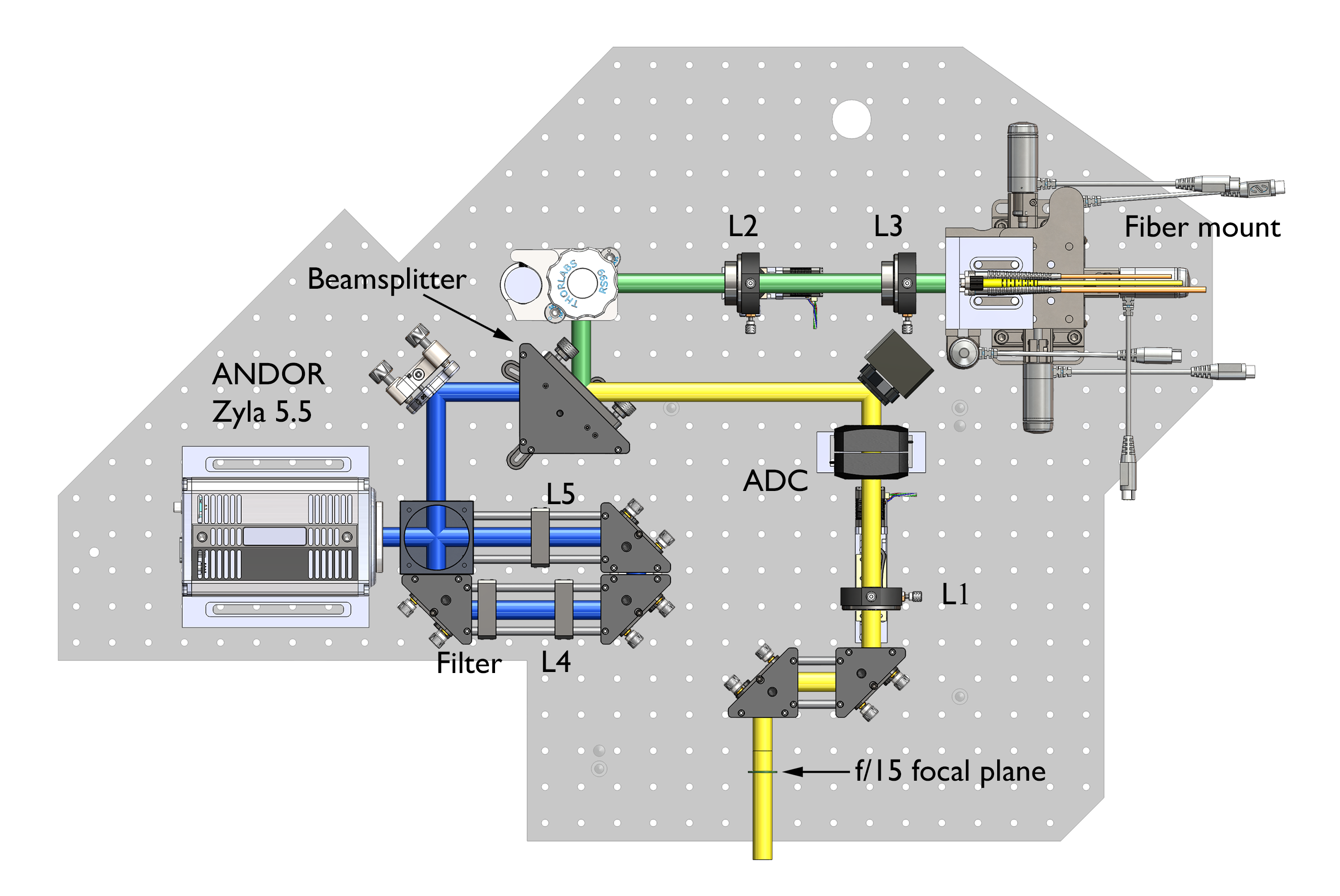}
   \end{tabular}
   \end{center}
   \caption{A simplified 3D CAD design of the AC demonstration system showing only the essential optical components. All wavelengths (yellow) propagate initially through the optical train. A beamsplitter directs Y- and J-band light toward the fiber arm (green). Shorter wavelengths are transmitted through the beamsplitter to the imaging arm.}
   { \label{fig:1} }
   \end{figure}

The schematic design of the AC system shown in Figure \ref{fig:1} can be divided into three sections: common optics (yellow), a fiber coupling arm (green), and an imaging arm (blue). The incident f/15 LBT beam enters at the bottom of the figure and is collimated using an f=200mm (Edmund Optics \#47-271) cemented achromatic lens (L1) to a diameter of 13.4mm before transmitting through an atmospheric dispersion corrector (ADC) composed of two singlet prisms (EKSMA 311-1200E). For efficient coupling, an ADC must be used to correct for dispersion created by propagation of starlight through the atmosphere \cite{Kopon}. A steering mirror (Optics In Motion OIM 5001) is used to correct for small angular deviations (up to 0.05$^{\circ}$) caused by the ADC as it compensates for different air masses. A laser line beamsplitter (Newport 10Q20HE.1) acts as the final common optic in the initial optical train, reflecting the science wavelength band to be used in the spectrograph to the fiber coupling arm of the instrument. The transmitted beam propagates through the beamsplitter into an imaging channel (blue) that is used to assist in fiber alignment, monitoring AO residuals, and tracking target star position.

\subsection{Fiber coupling arm} 
\label{subsec:Fiber coupling arm}
A single-mode fiber can only efficiently accept an optical beam with a spatial distribution that is approximately Gaussian\cite{Wagner}. The $1/e^{2}$ diameter\footnote{The $1/e^{2}$ diameter of a Gaussian function refers to the width at which the peak optical intensity has dropped to 13.5\%.} of a Gaussian function or mode field diameter (MFD), is used as a characteristic reference for spatial distribution matching between the injected beam and the optical fiber's accepted distribution. For optimal performance, the LBT PSF \footnote{Specifically the Airy disk core is approximated as a Gaussian function.} is re-imaged so that the $1/e^{2}$ diameter of the Airy core closely matches the MFD of the SMF across the desired coupling wavelength band. In order to avoid insertion loss, the geometric shape of the injected beam must also be within the numerical aperture (NA) accepted by the fiber. Matching the spatial distribution of the injected beam to the MFD of the fiber can be challenging for large wavelength bands as the injected PSF from a star increases (approximately) linearly with wavelength ($\lambda/D$) while the MFD of a fiber is approximated by the Petermann II mode field radius,

\begin{equation}
\label{eq:MFD}
\frac{\omega}{a} = 0.65 + \frac{1.619}{V^{3/2}} + \frac{2.879}{V^6} ,
\end{equation}
where $\omega$ represents the mode radius, $a$ is the physical fiber core diameter, and $V$ (known as the fiber $V$-number) is inversely proportional to wavelength.\cite{Petermann} 

   \begin{figure}[H]
   \begin{center}
   \begin{tabular}{c}
   \includegraphics[height=6cm]{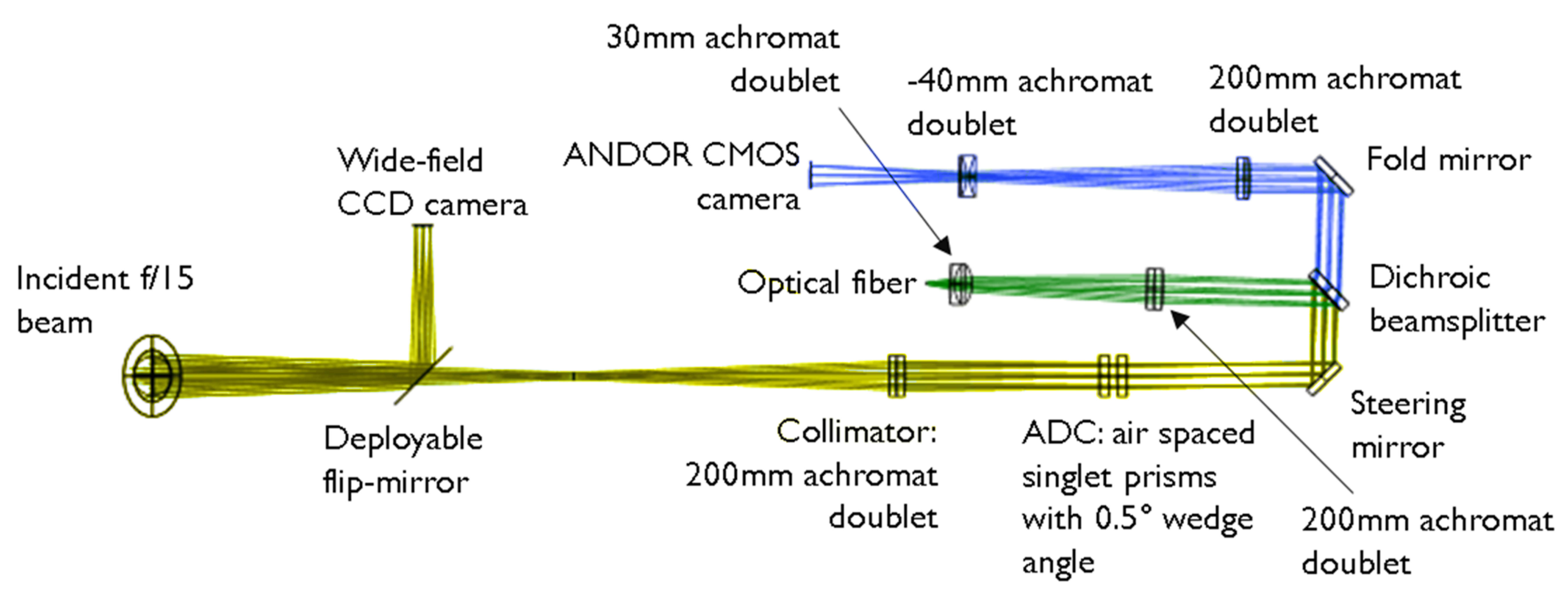}
   \end{tabular}
   \end{center}
   \caption{Zemax configuration of the AC demonstration system optical design. The fiber arm (green) uses a pair of focusing achromatic doublets to demagnify the LBT PSF to match the SMF's MFD for optimal coupling efficiency. Section \ref{subsec:Imaging arm} details the imaging arm (blue), which uses a pair of cemented achromatic lenses to magnify the LBT PSF to obtain an image of the target star with sufficient pixel sampling across the FWHM at 700nm.} 
   {\label{fig:2} 
}
   \end{figure} 

The fiber coupling arm of the instrument is designed to inject Y-band light from the telescope into a SMF with the proper optical characteristics required for optimal coupling efficiency. The PSF at the f/15 Gregorian focus of the LBT is much larger than the MFD of the fiber. As such, the optics in the acquisition camera fiber arm must demagnify the telescope beam to reduce the incoming $1/e^{2}$ width of the LBT PSF to closely match the fiber MFD (5.8$\mu m$ at 980nm) as well as the NA of the fiber (0.14). The separation of the fiber focusing optics can be altered by driving a linear stage (Micronix) to optimize the f-number for a given target star depending on the delivered AO correction. As shown in Figure \ref{fig:2}, the AC's fiber coupling arm focuses reflected science light (green) with a pair of achromatic lenses (Edmund Optics \#47-217, \#65-438), giving an effective focal length (EFL) of 48mm in order to match the injected beam width to the MFD over the full Y-band. A fiber stage (Newport 562 Series) driven by piezo actuators (Newport PZA12) is placed in the focal plane for precise ($<$0.1$\mu m$) positioning of the fiber. A custom fiber plate mounted on the stage can hold up to 3 optical fibers simultaneously (one SMF and two diagnostic MMFs). Diagnostic MMFs are included as a method of measuring the maximum power the optical system can couple to an optical fiber and to provide a fiber with a known location that is more easily coupled than the SMF. Switching between the MMFs and the SMF can be achieved by driving the fiber stage horizontally and vertically by a pre-measured number of microsteps. Optical encoders are used to ensure the absolute position of the fiber stage is known, as the piezo actuators have a very high precision and small incremental step (on average = 10nm) but low repeatability over the full range of motion of the stage.

Coupling efficiency (defined here as the amount of output power from the SMF in the LP$_{01}$ mode divided by the power of the incident telescope beam) will be measured with a power monitoring system comprising the three optical fibers, a one-to-one re-imaging system, optical filters, and an InGaAs sensor (Newport 818-IG). Incident power can be measured by deploying a mirror into the path of the injected beam and focusing the beam onto an identical photodiode sensor. Mauna Kea Y- (975-1075nm) and J-band (1160-1340nm) filters (Omega Optical) mounted in a filter wheel will be used to measure the power coupled in the respective bands throughout an integration. To switch between the three optical fibers in the output monitoring system, a selection mechanism consisting of a linear stage (Thorlabs Z825B) and custom optical fiber mount holding all three fibers can be used to reposition the selected fiber in the one-to-one re-imaging optics used for optical power measurement. Table 1 shows the optical characteristics of the three optical fibers that will be used in the on-sky experiment. 

\begin{table}
\caption{The AC demonstration system's optical fibers. The SM980-5.8-125 is the SMF which will be used for Y- and J-band coupling. The M65L01 and M14L01 are multi-mode fibers that are used both as alignment tools and to demonstrate the maximum power which can be coupled into conventional fibers.} 

\label{tab:fibers}
\begin{center}       
\begin{tabular}{|l|l|l|l|l|} 
\hline
\rule[-1ex]{0pt}{3.5ex}  Fiber Model Number & Vendor & MFD/Core Diameter($\mu$m)& NA & Cutoff Wavelength (nm) \\
\hline
\rule[-1ex]{0pt}{3.5ex}  SM980-5.8-125 & Fibercore & 5.8 at 980nm & 0.14 & 882   \\
\hline
\rule[-1ex]{0pt}{3.5ex}  M65L01 & Thorlabs & 10 & 0.10 &700-1400  \\
\hline
\rule[-1ex]{0pt}{3.5ex}  M14L01 & Thorlabs & 50 & 0.22 &700-1400  \\
\hline
\end{tabular}
\end{center}
\end{table} 

The theoretical coupling efficiency of a Gaussian intensity distribution to a SMF is near unity. However, when used in astronomical applications, the SMF will be coupled to an Airy pattern, whose near Gaussian core contains at most 78\%\cite{Shaklan}\footnote{without the use of corrective optics e.g.Phase Induced Amplitude Apodization (PIAA)\cite{Jovanovic}.} of the total power, which limits the theoretical maximum coupling efficiency to SMFs. Since previous data taken at the LBT indicates that a Strehl ratio of 60\% is possible in the Y-band\cite{IRTC}, and the maximum achievable coupling efficiency to SMFs can be estimated by the Strehl ratio\cite{Shaklan}, the LBT AO correction provides a promising site for efficient SMF coupling.   

\subsection{Imaging Arm} 
\label{subsec:Imaging arm}
The imaging arm of the instrument receives wavelengths shorter than 950nm via transmission through the beamsplitter. Due to the 6.35mm thickness of the beamsplitter optic, a lateral deviation of 3.5mm from the optical axis is introduced and is compensated for in the optical alignment. A small wedge on the back surface ($<$3') causes minimal dispersion and angular deviation. A long-pass filter (Thorlabs FELH700) used in combination with the beamsplitter creates an effective a bandpass of 700-950nm in the imaging camera. 

A pair of cemented achromatic doublets (Thorlabs AC254-200-B, ACN254-040-B) with an EFL of 750mm focus the transmitted beam onto an ANDOR Zyla 5.5 CMOS detector (6.5$\mu m$ pixel pitch), giving a pixel scale of 2.88 mas/pixel. This value exceeds iLocater's requirement of at least 5 pixels across the FWHM for sufficient PSF sampling at the minimum wavelength while simultaneously providing a field of view (FOV) of at least 4$\times$4'' across the 2516$\times$2160 pixel array. 

Images taken during the fiber coupling process will provide pseudo real-time feedback to assist with alignment of the telescope axis to the optical axis of the fiber. Predefined pixel coordinates corresponding to the location of the fiber in the ANDOR focal plane will act as a default fiducial reference for centering the star. Once the telescope is aligned to the instrument, monitoring of the  centroid  of the telescope PSF can be used to reposition the optical beam onto the fiber pixel coordinates. Frames will be recorded throughout on-sky commissioning as a reference for AO residuals and Strehl ratio estimation. 

\section{MECHANICAL DESIGN}
\label{sec:MECHANICAL DESIGN}
\begin{figure}[H]
   \begin{center}
   \begin{tabular}{c}
   \includegraphics[height=7cm]{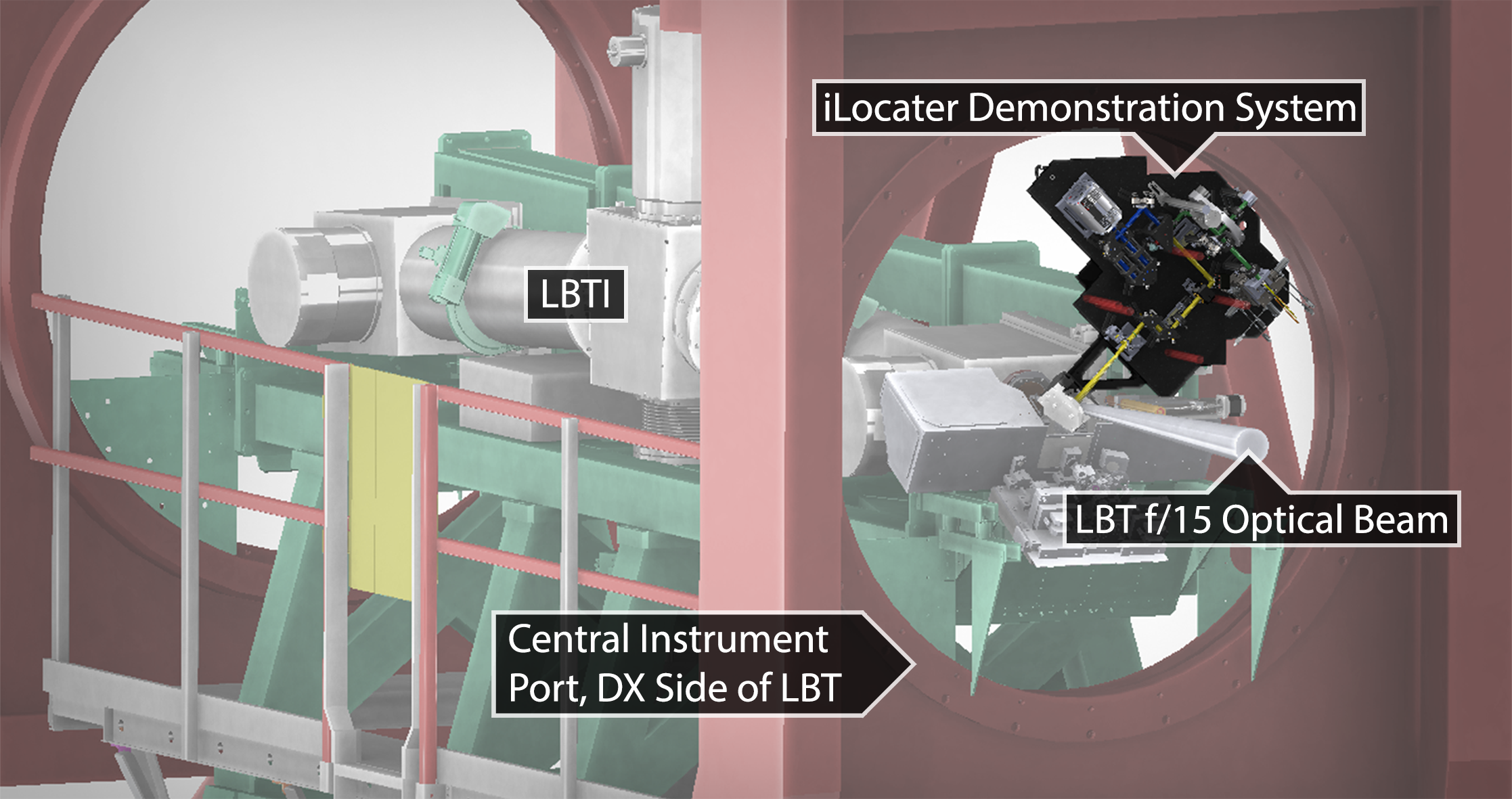}
   \end{tabular}
   \end{center}
   \caption{A 3D CAD rendering showing the location of the AC demonstration system as mounted on LBTI. The instrument is located within the central LBT instrument port structure shown in red. Incident light from the LBT enters the port and is diverted toward the AC optics by a 50/50 beamsplitter and fold mirror in a pick-off arm installed by the INAF SHARK team.} 
   { \label{fig:4} 
 }
   \end{figure}

The iLocater demonstration system will be installed on the DX side of the LBT, positioned directly after the central bent-Gregorian instrument port. The INAF SHARK instrument team\footnote{Instituto Nazionale di Astrofisica (INAF), \textbf{S}ystem for Coronography with \textbf{H}igh Order \textbf{A}daptive Optics from \textbf{R} to \textbf{K} Band (SHARK).} have designed and installed a mounting structure for a modestly sized (425$\times$425mm) optical board. iLocater reuses the same mounting structure to install a custom optical board (800$\times$500MM) for commissioning the demonstration system. As shown in Figure \ref{fig:4}, the optical board will be oriented vertically, such that is co-rotates with the telescope (altitude) within the instrument port. The changing gravity vector between Zenith and lowest pointing above the horizon (90 - 20$^{\circ}$) makes the stability of the optical mounts essential. All stages needed to perform under these conditions have been tested in the lab under various orientations to check repeatability before use in the demonstration system. Once the custom optical board is installed, access to the optics is limited due to the mounting location and orientation (towards the primary mirror) as well as the instrument enclosure requiring many of the critical alignment components to be remotely controlled. 

\subsection{Full Demonstration System layout} 
\label{subsec:Full instrument layount}
\begin{figure}
   \begin{center}
   \begin{tabular}{c}
   \includegraphics[height=7cm]{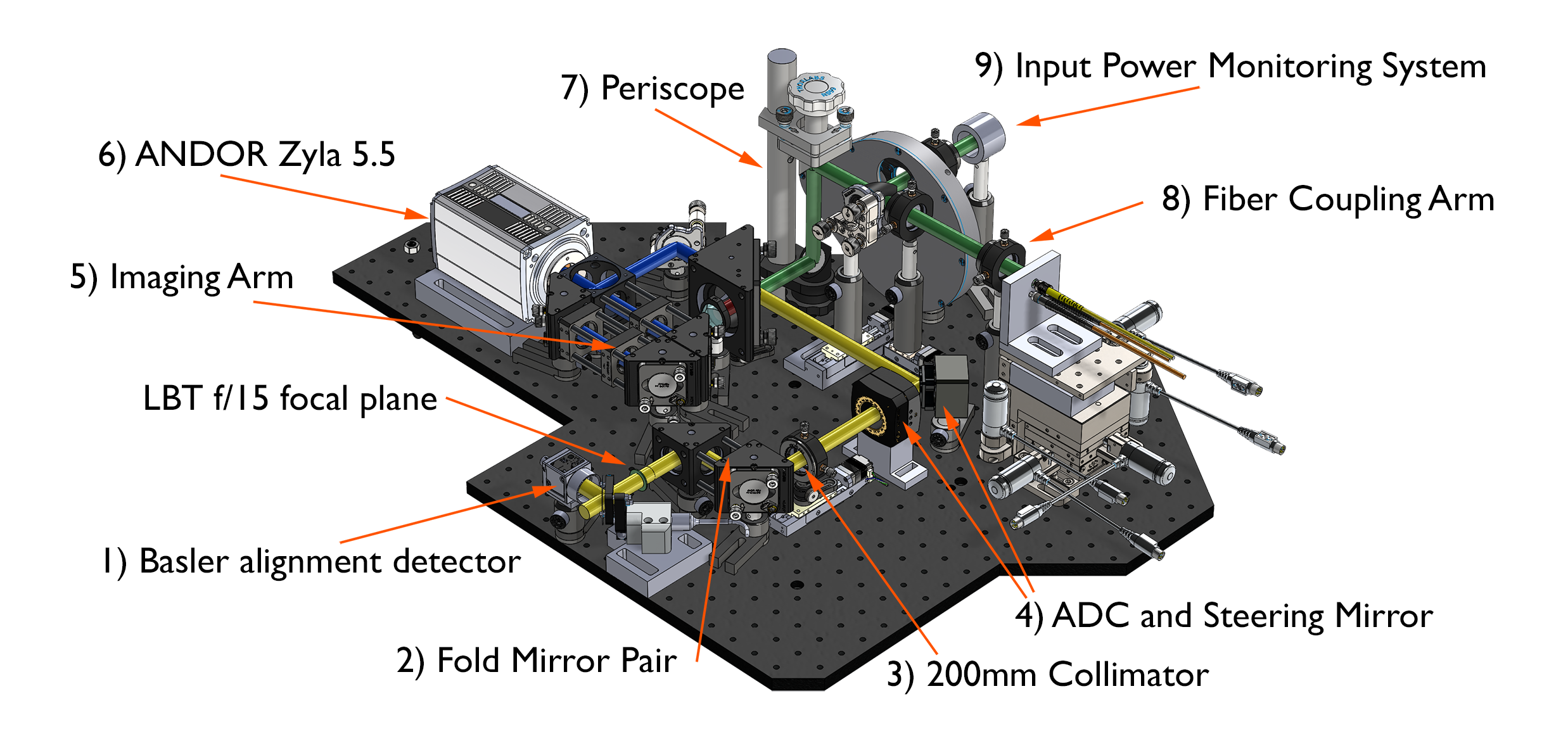}
   \end{tabular}
   \end{center}
   \caption{A 3D CAD rendering showing the full instrument design. Significant folding of the optical beam is required to fit the system onto the compact optical board. The full beam path is shown, including all optical components and optomechnics with combined optics in yellow, fiber coupling arm optics in green, and imaging optics in blue.}
   { \label{fig:5} }
   \end{figure}

The full AC layout includes additional components previously not shown in the optical design. These include devices for optical power monitoring, steering of the optical beam, and folding to keep the beam within the compact envelope of the optical board. The full optical layout is shown in Figure \ref{fig:5}. 

A CMOS (Basler acA2040-90um) alignment camera with a 10$\times$10'' FOV is located at the entrance of the instrument (1) and can be used for coarse alignment of the target star to the instrument optical axis. Light can be diverted to this detector by deploying a 25mm diameter flip mirror. Cage mounted fold mirrors (2) with tip/tilt adjustment redirect the f/15 beam toward the collimator lens (3). All fold mirrors have a protected gold coating from Newport with $\lambda /20$ surface accuracy and reflectance $>97 \%$ at NIR wavelengths. The ADC prism pair (4) is used in collimated space to correct for dispersion. A steering mirror located immediately after the ADC can be used for fine alignment of the target star in the imaging channel (5) on the ANDOR CMOS detector(6). The fiber coupling arm (8) is located after a periscope (7) that raises the beam height to the correct height for the fiber stage. Input power is measured by deploying a 25mm mirror into the fiber arm immediately after the periscope (9). As explained in Section \ref{subsec:Fiber coupling arm}, a photodiode sensor will monitor the input power in the Y-band or J-band by rotation of a filter wheel and used for fiber coupling efficiency calculations.

\subsection{Data Recording} 
\label{subsec:Recording data}
Power meter data will be recorded using two identical InGaAs photodiodes programmed at the center wavelengths of the Y- and J-bands (1025 and 1250nm) respectively. In addition, the demonstration system will be used as a diagnostic testbed capable of measuring and recording environmental (vibrational, thermal) and optical stability (wavefront residuals, pointing) data. These measurements can be used to identify systematic effects seen in fiber coupling and will provide essential data for improving the design of iLocater's final acquisition camera. 

Real-time vibrational data will be recorded by mounting a three-axis accelerometer to the optical board. Vibrations at the SHARK board location on LBTI will be compared against the independent OVMS\footnote{Optical path difference and Vibration Monitoring System (OVMS)} accelerometer data recorded within the central LBTI green structure (Figure \ref{fig:4}). Thermal measurements will be simultaneously logged at the five-axis fiber stage and near the edge of the instrument enclosure using two J-type exposed-junction thermocouples (National Instruments 781314-02). The ANDOR Zyla 5.5 CMOS can record images of the target PSF at frame rates up to f=1kHz, which will be stored on a separate computer with RAID setup to rapidly store images. 


\section{PRELIMINARY LAB RESULTS}
\label{sec:LAB RESULTS}
Initial tests of the demonstration system are currently being undertaken in the Astrophysics lab at the University of Notre Dame. An LBT simulator has been constructed using OAPs to output an f/15 beam with no chromatic aberrations. The optical output of this system matches the f-number of the LBT to within 2$\%$ and PSF diameter at the focal plane to within 1$\mu m$ of the theoretical telescope value (24$\mu m$ $1/e^{2}$ diameter at 980nm). A custom focal plane mask machined in-house reproduces the secondary mirror and spider obscurations to replicate similar diffraction effects and the correct spatial resolution of the LBT support structure. For fiber coupling measurements, a super-continuum (fianium WhiteLase micro) source is used to inject bright broadband light into the LBT simulator system while a 635nm or 1064nm benchtop fiber coupled laser can be used for alignment. 

  \begin{figure}[h]
  \begin{center}
  \begin{tabular}{c}
  \includegraphics[height=7cm]{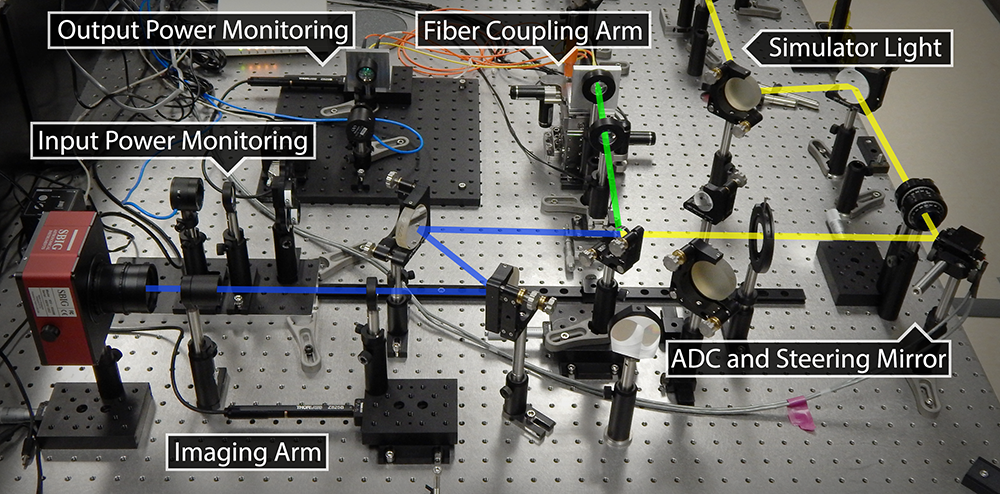}
  \end{tabular}
  \end{center}
  \caption{Photograph of the demonstration system test bed at the University of Notre Dame. Light enters from the top of the system. The fiber coupling arm is used to test specific optics and fiber optic performance in the Y-band. In addition to fiber coupling, this system is used for fiber alignment techniques in conjunction with the imaging arm, measuring the repeatability and precision of motion stages, and calibrating power monitoring systems.}
  {\label{fig:6} 
 
}
  \end{figure} 

The LBT simulator system injects light into an unfolded version of the demonstrator system with equivalent optical specifications as the AC in Figure \ref{fig:6}. The testbed system allows broadband experiments of all optics and testing of selected opto-mechanics prior to use on-sky. Current lab results demonstrate routine SMF coupling efficiency $>$70\% over the full Y-band. A lower coupling efficiency is expected (60\%) in the J-band due to a slight mismatch of the injected PSF compared to the SM980 fiber MFD at 1250nm as explained in Section \ref{subsec:Fiber coupling arm}. Using filters with 10nm FWHM bandwidth, preliminary measurements in the Y-band show the light coupled to the SM980 fiber is distributed equally over 975-1075nm. 

\section{CONCLUSIONS}
\label{sec:CONCLUSIONS}
Recent advances in adaptive optics technology make it possible to deliver a stable, diffraction-limited beam to instruments operating at NIR wavelengths. In principle, this achievement allows for the efficient coupling of starlight into single-mode optical fibers for use in a new generation of fiber-fed, diffraction-limited RV spectrographs that will overcome the current limitations set by the effects of modal noise. Scheduled for commissioning at the LBT this fall, the iLocater AC demonstration system will be the first instrument to directly measure SMF coupling efficiencies on-sky using an ``extreme" adaptive optics system on an 8-10 meter class telescope. In addition to recording relative output power of a SMF in the Y-band, the demonstrator system will also serve as a useful diagnostic system that informs iLocater's final acquisition camera design by correlating fiber insertion loss with vibrations, thermal effects, and Strehl ratio fluctuations. 

\acknowledgments     
iLocater's acquisition camera is funded in part through PI. Crepp's NASA Early Career Award. The iLocater team is also deeply grateful for support from the Potenziani family for their generosity and vision. 

\bibliography{biblist}   
\bibliographystyle{spiebib}   

\end{document}